\documentclass[11pt]{article}
\newcommand{\be}{\begin{equation}}
\newcommand{\ee}{\end{equation}}
\newcommand{\bea}{\begin{eqnarray}}
\newcommand{\eea}{\end{eqnarray}}

\newcommand{\sn}{{\rm sn}}
\newcommand{\cn}{{\rm cn}}
\newcommand{\dn}{{\rm dn}}

\begin{document}
~\hfill{\footnotesize SUNYB/05-13, IOPB/2005-07}
\vspace{.5in}
\begin{center}
{\LARGE {\bf Quasi-Periodic Solutions of Heun's Equation}} 
\end{center}
\vspace{.3in}
\begin{center}
{\Large{\bf  \mbox{Avinash Khare}}}\\
\noindent
{\large Institute of Physics, Sachivalaya Marg, Bhubaneswar 751005, India}
\end{center}
\begin{center}
{\Large{\bf  \mbox{Uday Sukhatme}}}\\
\noindent
{\large Department of Physics, State University of New York at Buffalo, Buffalo, NY 14260, U.S.A. }\\
\end{center}
\vspace{.9in}
{\bf {Abstract:}}  
By exploiting a recently developed connection between Heun's differential equation and the generalized
associated Lam\'e equation, we not only recover the well known periodic solutions, but also obtain a 
large class of new, quasi-periodic solutions of Heun's equation. Each of
the quasi-periodic solutions is doubly degenerate.
\newpage
Heun's equation, a second order linear differential
equation with four regular singular points has been extensively
discussed in the mathematics literature \cite{ron,mai,erd}. 
In recent years, this equation has also appeared in a number of
physical problems, like quasi-exactly solvable systems \cite{jnk},
sphaleron stability \cite{bre}, Calogero-Sutherland models \cite{tak},
higher dimensional correlated systems \cite{bklms}, Kerr-de Sitter black
holes \cite{stu}, and finite lattice Bethe ansatz systems \cite{dst}.

The so-called periodic (also termed as polynomial) solutions 
of Heun's equation have been well studied. However, as emphasized 
in ref. \cite{ron}, much less attention has been devoted to the quasi-periodic (also termed as
non-polynomial) solutions. In this letter we obtain a large class of
(mostly new) quasi-periodic solutions. We shall show that each such solution is doubly degenerate.

The canonical form of Heun's equation is given by \cite{ron}
\be\label{1}
\bigg [\frac{d^2}{dx^2}+\bigg (\frac{\gamma}{x}+\frac{\delta}{x-1}
+\frac{\epsilon}{x-c} \bigg )\frac{d}{dx}+\frac{\alpha \beta
x-q}{x(x-1)(x-c)} \bigg ]G(x) =0\,,
\ee
where $\alpha,\beta,\gamma,\delta,\epsilon,q,c$ are parameters. 
The parameters are not all independent; the constraint relation is 
\be\label{1a}
\gamma+\delta+\epsilon=\alpha+\beta+1\,.
\ee
The four regular singular points of the differential equation are 
located at $x=0, \,1, \,c ~(\ne 0,1)$, and the point at infinity. 

Let us make a change of independent variables using the transformation
$x=\sn^2(y,m)$ \cite{abr}, where $m \equiv 1/c$. Then Heun's equation takes the form \cite{ron} 
\bea\label{2}
&&F''(y)+ \bigg [(1-2\epsilon) m \frac{\sn(y,m) \cn(y,m)}{\dn(y,m)}
+(1-2\delta)
\frac{\sn(y,m)\dn(y,m)}{\cn(y,m)} +(2\gamma-1) \frac{\cn(y,m)
\dn(y,m)}{\sn(y,m)}\bigg ]F'(y) \nonumber \\
&&~~~~~~~~~~~~~~~~~~~~~~~~~~~~~~~~~~-[4mq-4\alpha \beta m \sn^2 (y,m)]F(y) =0\,,
\eea
where $G(x) \equiv F(y)$.
The periodic solutions of eq. (\ref{2}) correspond to the 
polynomial solutions of eq. (\ref{1}) while the quasi-periodic solutions
of this equation correspond to the non-polynomial solutions of eq. (\ref{1}).

Before describing the quasi-periodic solutions of eq.
(\ref{2}), it may be worthwhile explaining how we arrived at such solutions.
A few years ago, we studied in some detail the Schr\"odinger equation 
for the associated Lam\'e
(AL) potentials \cite{ks1,ks2}
\be\label{3}
V(y,m)=a(a+1)m\sn^2(y,m)+b(b+1)m{\sn^2 (y+K(m),m)}=a(a+1)m \sn^2(y,m) +b(b+1)m \frac{\cn^2(y,m)}{\dn^2(y,m)}\,,
\ee
which after a transformation can be shown to be a special case of eq.
(\ref{2}).
In particular, the band edges and the mid-band states of several of
these potentials were studied, when $a$ and $b$ were related in various specific ways. 
Further, when $a,b$ are both integers, we showed that these
potentials had the special feature of possessing only a finite number of band gaps.
This study has been extended to the generalized associated Lam\'e (GAL) potentials \cite{tv,ks3}
\bea\label{4} 
V(y,m)&&\!\!\!\!\!\!\!\!\!\!=a(a+1)m\sn^2(y,m)+b(b+1)m{\sn^2 (y+K(m),m)} \nonumber \\
&&~~~~~+f(f+1)m {\sn^2 (y+K(m)+iK'(m),m)}+g(g+1)m{\sn^2 (y+iK'(m),m)} \nonumber \\
&&\!\!\!\!\!\!\!\!\!\!= a(a+1)m\sn^2(y,m)+b(b+1)m\frac{\cn^2 (y,m)}{\dn^2 (y,m)} 
+f(f+1) \frac{\dn^2 (y,m)}{\cn^2(y,m)} 
+g(g+1)\frac{1}{\sn^2 (y,m)}~. \nonumber \\
\eea
The point to note is that after a transformation, the
Schr\"odinger equation for the potential given in eq. (\ref{4}) 
is identical in form to eq.
(\ref{2}). In particular, let us start from the Schr\"odinger equation
\be\label{5}
-\frac{d^2 \psi(y)}{dy^2}+V(y,m)\psi(y)=E\psi(y)\,,
\ee
with $V(y,m)$ given by eq. (\ref{4}). On substituting
\be\label{6}
\psi(y)=\dn^{-b}(y)\cn^{-f}(y)\sn^{-g}(y)\phi(y)\,,
\ee
one can show that $\phi(y)$ satisfies Heun's equation
(\ref{2}). More precisely, $\phi(y)$ satisfies the differential equation
\be\label{7}
\phi''(y)+2 \bigg[ mb \frac{\sn(y,m) \cn(y,m)}{\dn(y,m)}
+f\frac{\sn(y,m)\dn(y,m)}{\cn(y,m)} -g \frac{\cn(y,m)
\dn(y,m)}{\sn(y,m)} \bigg ]\phi'(y)-[R
-Q m \sn^2 (y,m)]\phi(y) =0\,,
\ee
where 
\be\label{8}
R=-E+m(g+b)^2 +(f+g)^2\,,~~Q=(b+f+g)(b+f+g-1)-a(a+1)\,.
\ee
Thus, once we obtain solutions of the Schr\"odinger equation
for the GAL potential (\ref{4}), then we can immediately write the
solutions of eq. (\ref{2}) and hence that
of the original Heun's eq. (\ref{1}) with the identification
\bea\label{9}
&&\gamma = \frac{1}{2} -g\,,~~\delta = \frac{1}{2} -f\,,~~\epsilon =
\frac{1}{2}-b\,, ~~\alpha+\beta =\frac{1}{2}-(b+f+g)\,,\nonumber \\
&&~4\alpha \beta =Q\,,~~4mq=R\,,
~~F(y) \equiv \phi(y)\,.
\eea

A few comments are in order.
\begin{enumerate}

\item The Schr\"odinger eq. (\ref{5}) for the 
GAL potential (\ref{4}) is invariant
under $y \rightarrow y+K(m)$ provided $a$ and $b$ are interchanged and
so also are $f$ and $g$. Hence, the eigenvalues $E$ are invariant under 
$a \leftrightarrow b$ and $f \leftrightarrow g$
while the corresponding eigenfunctions are
related to each other by the translation $y \rightarrow y+K(m)$.
Likewise, the eigenvalues $E$ 
are also invariant under 
$a \leftrightarrow f$ and $b \leftrightarrow g$,
while the corresponding eigenfunctions are
related to each other by the translation $y \rightarrow
y+K(m)+iK'(m)$. Similarly, the eigenvalues $E$ are invariant under the
transformation $a \leftrightarrow g$ and $b \leftrightarrow f$, 
while the corresponding eigenfunctions are
related to each other by the translation $y \rightarrow y+iK'(m)$.

\item The Schr\"odinger eq. (\ref{5}) for the GAL potential
(\ref{4}) is also invariant under $a \rightarrow -a-1$ and/or $b
\rightarrow -b-1$ and/or $f \rightarrow -f-1$ and/or $g \rightarrow 
-g-1$. As a result the eigenvalues $E$ as well as the
corresponding eigenfunctions are invariant under one or several of these
transformations.

\item It may be noted that, except for the invariance under 
$a \rightarrow -a-1$, eq. (\ref{2}) is not
invariant under any of the above transformations. However, the
connection between the GAL problem and eq.
(\ref{2}) and the invariances of the GAL
equation, can be exploited to obtain several more solutions of Heun's
equation. For example, from eq. (\ref{8}), it follows that if under
any of the above transformations, if $b_1,f_1,g_1$ change to
$b_2,f_2,g_2$ and the energies $E$ 
remain invariant, then  
the corresponding eigenvalues $R=4mq$ of Heun's eq. (\ref{2})
 are related by
\be\label{9a}
R_1 -m(b+1+g_1)^2-m(f_1+g_1)^2=R_2 -m(b_2+g_2)^2-(f_2+g_2)^2\,.  
\ee
\end{enumerate}

Our strategy is now clear. We shall first obtain solutions of the
Schr\"odinger eq. (\ref{5}) for the GAL potential (\ref{4})
and then using the connections given in eqs. (\ref{9}) and
(\ref{9a}) and the symmetries of the
GAL equation, we shall obtain a host of solutions of 
Heun's equation. This strategy is demonstrated below by discussing one
example in detail. As mentioned previously, the focus here is on the quasi-periodic
solutions of the GAL equation.

We now show that when either $a+b+f+g$ and/or $a-b-f-g$
is an arbitrary half-integer (but not an integer), 
then one can obtain doubly degenerate eigenstates of the 
GAL eq. (\ref{5})
which correspond to the mid-band states (rather than band-edges) 
of this periodic problem. 
In the special case when $a+b+f+g$ and/or $a-b-f-g$ is an integer,
then one obtains two distinct (non-degenerate) eigenstates which
correspond to the band edge eigenstates. It turns out that   
depending on whether 
$b$ or $f$ or $g$ is half-integral (while the other two are integral), 
we need to use a different ansatz. Let us consider all three cases one
by one.

{\bf Case 1: $b$ half-integral}

We start from eq. (\ref{7}) and substitute the ansatz
\be\label{10}
\phi(y) = [\cn(y,m)+i\sn(y,m)]^{t} Z(y)\,,
\ee
where $t$ is any real number.
It follows that $Z(y)$ satisfies the equation
\bea\label{11}
&&Z''(y)+[2it\dn(y,m)+2mb\frac{\sn(y,m)\cn(y,m)}{\dn(y,m)}
-2g\frac{\cn(y,m)\dn(y,m)}{\sn(y,m)}+2f\frac{\dn(y,m)\sn(y,m)}
{\cn(y,m)}]Z'(y) \nonumber \\
&&+[-(R+t^2)+(Q+t^2)m\sn^2(y,m)-2itg\frac{\cn(y,m)}{\sn(y,m)}
+2itf(1-m)\frac{\sn(y,m)}{\cn(y,m)} \nonumber \\
&&+imt(2b+2f+2g-1)\sn(y,m)\cn(y,m)]Z(y)=0\,,
\eea 
where $R$ and $Q$ are as given by eq. (\ref{8}).
Not surprisingly, $Z(y)=$ constant is a solution with energy
$E=(4t^2+m)/4$ provided $f=g=0,~b=1/2,~a=t-1/2$ (i.e. $b+f+g=1/2$).  

One can build solutions for higher half-integer values of $b+f+g$ from here. In
particular, if $b+f+g=2M+\frac{1}{2}$, we choose the ansatz
($M=0,1,2,...$)
\be\label{12}
Z(y)=\sum_{k=0}^{M} A_k\,\, \sn^{2k}
(y,m)+\cn(y,m)\sn(y,m)\sum_{k=0}^{M-1} B_k\, \sn^{2k} (y,m)\,,
\ee
while if $b+f+g=2M+3/2,$ then we consider the ansatz ($M=0,1,2,...$)
\be\label{13}
Z(y)=\cn(y,m)\sum_{k=0}^{M} A_k \sn^{2k}
(y,m)+\sn(y,m)\sum_{k=0}^{M} B_k \sn^{2k} (y,m)\,.
\ee
Substitution into eq. (\ref{11}) followed by lengthy but
straightforward algebra yields analytic expressions for the
energy eigenvalues  for arbitrary $M$ and
$b=1/2,3/2.$ In particular, for $b=1/2$, we find
\be\label{14}
b=1/2\,,~ f+g=N\,,~g=p\,,~ a=t-1/2\,,~~E=[t^2+m(g+b)^2]\,, 
\ee
where both $f,g$ are nonnegative integers satisfying $f+g=N$ with
$p,N=0,1,2,...$ . Since $p\le N$, it follows that for a given $N$,
there are $N+1$ different solutions.
Similarly, when $b=3/2$, we get $g=p,f=N-p,a=t-1/2$, and
\be\label{15}
E=[1+t^2+m(g+b)^2]-m(2g+1) \pm \sqrt{(2g+1)^2m^2
+4m(N+1)(f-g)+4(1-m)t^2}\,
\ee
where, $f$ and $g$ are again nonnegative integers. Note that, in this case
too, for a given $N$, there are $N+1$ different solutions. 
In all cases, the
corresponding eigenfunctions have the form given in eqs.
(\ref{12}) and (\ref{13}). For small
 values of $N$, the explicit coefficients $A_k,B_k$ appearing in the
eigenfunction expressions can be easily written
down. 

We can immediately write down the solutions of Heun's eq.
(\ref{2}) by making use of eq. (\ref{9}). 
In particular, we find the following two classes of solutions:

\be\label{16}
\gamma = \frac{1}{2}-p\,,~\delta=\frac{1}{2}-N+p\,,~\epsilon=0\,,
\alpha=-\frac{N-t}{2}\,,~ \beta=-\frac{N+t}{2}\,,~4mq=N^2-t^2\,;
\ee
\bea\label{17}
&&\gamma = \frac{1}{2}-p\,,~\delta=\frac{1}{2}-N+p\,,~\epsilon=-1\,,
\alpha=-\frac{N+1-t}{2}\,,~ \beta=-\frac{N+1+t}{2}\,, \nonumber \\
&&4mq=N^2-t^2-1+(2p+1)m \pm \sqrt{(2p+1)^2 m^2
+4m(N+1)(N-2p)+4(1-m)t^2}\,.
\eea
The corresponding eigenfunctions are generically as given by 
eqs. (\ref{10}), (\ref{12}) and (\ref{13}) with $F(y) \equiv \phi(y)$.
As an illustration, for $N=1$, the eigenfunction is $Z(y)=
A\cn(y,m)+B\sn(y,m)$ with $\frac{B}{A}=it$ for $f=1,g=0,b=1/2$, 
while for $g=1,f=0,b=1/2$, one gets $\frac{B}{A}=i$.

Several remarks are in order at this stage.

\begin{enumerate}

\item Since the Schr\"odinger eq. (\ref{5}) for the GAL potential 
(\ref{4}) as well as Heun's eq. (\ref{2}) are invariant under $y
\rightarrow y+2K(m) and y \rightarrow y+2iK'(m)$, 
hence $F(y)$, $F(y+2K(m))$ and 
$F(y+2iK'(m))$ are all eigenfunctions of Heun's eq. (\ref{2}) 
with the same eigenvalue. As a consequence,  
$F(y)=[\cn(y,m)-i\sn(y,m)]^{t}Z(y)$ is also an eigenfunction
of Heun's eq. (\ref{2}) with  
the {\it same} eigenvalue. Thus for any nonintegral $t$, each level is
doubly degenerate. The same remark also applies to the other two
solutions (when $f$ or $g$ are half integral) discussed below.

\item For the special case of $f=g=0$ and $t=1/2$, the results (as
expected) are identical to those already
obtained in ref. \cite{ks2}.   

\item For integral $t$, both $a,b$ are half integral (while $f,g$ are 
integral) and each of these 
solutions reduces to two nondegenerate periodic solutions of the GAL equation. 

\item We can generate new solutions of Heun's eq. (\ref{2}) by 
considering the transformations $y \rightarrow
y+K(m),y \rightarrow y+iK'(m),y \rightarrow y+K(m)+iK'(m)$ and the
fact that the eigenvalues are
invariant provided we interchange $a,b,f,g$ appropriately as
explained above. Obviously, each of these new solutions is also
doubly degenerate. In particular, by starting from solution
(\ref{16}) and using eq. (\ref{9a}), 
the three sets of new solutions of Heun's eq. (\ref{2}) are
\bea\label{18}
&&\gamma = \frac{1}{2}-N+p\,,~\delta = \frac{1}{2}-p\,,~\epsilon=1-t\,,
\alpha=-\frac{N+t-2}{2}\,, \nonumber \\
&&\beta=-\frac{N+t}{2}\,,~4mq=N^2-t^2+m(N+t)[N+t-2p-1]\,,
~F \equiv F(y+K(m))\,.
\eea
\bea\label{19}
&&\gamma = 1-t\,,~\delta =0\,,~\epsilon=\frac{1}{2}-N+p\,,
\alpha=-\frac{N+t-p-1}{2}\,, \nonumber \\
&&\beta=-\frac{N+t}{2}\,,~4mq=m(N+t)[N+t-2p-1]\,, ~F \equiv
F(y+iK'(m))\,.
\eea
\bea\label{20}
&&\gamma=0\,,~\delta=1-t\,,~\epsilon = \frac{1}{2}-p\,,~
\alpha=-\frac{p+t-1}{2}\,, \nonumber \\
&& \beta=-\frac{p+t}{2}\,,~4mq=0\,, ~F \equiv F(y+K(m)+iK'(m))\,.
\eea
Similarly, by starting from solution (\ref{17}) with $b=3/2$, 
three sets of new solutions can be immediately written down.

\item For low values of $N$, the corresponding eigenfunctions can be
explicitly shown. For example, for $g=0,f=1$, i.e. $p=0,N=1$, the
eigenfunction corresponding to solution (\ref{16}) is given by
\be\label{21}
F(y)=[\cn(y,m)+i\sn(y,m)]^{t} [\cn(y,m)+it\sn(y,m)]\,,
\ee
while the eigenfunctions corresponding to solutions (\ref{18}) to
(\ref{20}) are given by
\be\label{22}
F \equiv F(y+K(m) ) \propto 
[\cn(y,m)+i\sqrt{1-m}\sn(y,m)]^{t} [t\cn(y,m)+i\sqrt{1-m}\sn(y,m)]
[\dn(y,m)]^{-(1+t)}\,,
\ee
\be\label{23}
F \equiv F(y+iK'(m) ) \propto 
[1-\dn(y,m)]^{t} [t-\dn(y,m)]
[\sn(y,m)]^{-(1+t)}\,,
\ee
\be\label{24}
F \equiv F(y+K(m)+iK'(m) ) \propto 
[\dn(y,m)-\sqrt{1-m}]^{t} [t\dn(y,m)-\sqrt{1-m}]
[\cn(y,m)]^{-(1+t)}\,.
\ee

\item We can generate even more new solutions of Heun's eq. (\ref{2})
by considering the transformations $b \rightarrow -b-1$, $f
\rightarrow -f-1$ and $g \rightarrow -g-1$ either singly or in various
combinations. As an illustration, by starting from solution (\ref{16})
and using eq. (\ref{9a}), we get the following seven sets of new 
solutions.
\bea\label{24a}
&&\gamma=\frac{1}{2}-p\,,~\delta =\frac{1}{2}+p-N\,,~\epsilon=2\,,~
\alpha=-\frac{N-t-2}{2}\,, \nonumber \\
&&\beta=-\frac{N+t-2}{2}\,,~4mq=N^2-t^2-2m(2p-1)\,,
~F \equiv \frac{F(y)}{\dn^2(y)}\,.
\eea
\bea\label{24b}
&&\gamma=\frac{1}{2}-p\,,~\delta =\frac{3}{2}+N-p\,,~\epsilon=0\,,~
\alpha=\frac{N+t+1-2p}{2}\,, \nonumber \\
&&\beta=-\frac{N+1-t-2p}{2}\,,~4mq=N^2-t^2-(2p-1)(2N-2p+1)\,,
~F \equiv \frac{F(y)}{\cn^{2N-2p+1}(y)}\,.
\eea
\bea\label{24c}
&&\gamma=\frac{3}{2}+p\,,~\delta =\frac{1}{2}+p-N\,,~\epsilon=0\,,~
\alpha=\frac{2p+t+1-N}{2}\,, \nonumber \\
&&\beta=-\frac{2p+1-t-N}{2}\,,~4mq=N^2-t^2-(2p+1)(2N-2p-1)\,,
~F \equiv \frac{F(y)}{\sn^{2p+1}(y)}\,.
\eea
\bea\label{24d}
&&\gamma=\frac{1}{2}-p\,,~\delta =\frac{3}{2}+N-p\,,~\epsilon=2\,,~
\alpha=\frac{N+t+3-2p}{2}\,,~
\beta=-\frac{N+3-t-2p}{2}\,, \nonumber \\
&&4mq=N^2-t^2-2m(2p-1)-(2p-1)(2N-2p+1)\,,
~F \equiv \frac{F(y)}{\dn^2(y)\cn^{2N-2p+1}(y)}\,.
\eea
\bea\label{24e}
&&\gamma=\frac{3}{2}+p\,,~\delta =\frac{1}{2}+p-N\,,~\epsilon=2\,,~
\alpha=\frac{2p+t+3-N}{2}\,,~
\beta=-\frac{2p+3-t-N}{2}\,, \nonumber \\
&&4mq=N^2-t^2+2m(2p+3)-(2p+1)(2N-2p-1)\,,
~F \equiv \frac{F(y)}{\dn^2(y)\sn^{2p+1}(y)}\,.
\eea
\bea\label{24f}
&&\gamma=\frac{3}{2}+p\,,~\delta =\frac{3}{2}+N-p\,,~\epsilon=0\,,~
\alpha=\frac{N+t+2}{2}\,,~ 
\beta=-\frac{N+2-t}{2}\,, \nonumber \\
&&4mq=N^2-t^2+4(N+1)\,,
~F \equiv \frac{F(y)}{\cn^{2N-2p+1}(y)\sn^{2p+1}(y)}\,.
\eea
\bea\label{24g}
&&\gamma=\frac{3}{2}+p\,,~\delta =\frac{3}{2}+N-p\,,~\epsilon=2\,,~
\alpha=\frac{N+t+4}{2}\,,~ 
\beta=-\frac{N+4-t}{2}\,, \nonumber \\
&&4mq=N^2-t^2+2m(2p+3)+4(N+1)\,,
~F \equiv \frac{F(y)}{\dn^2(y)\cn^{2N-2p+1}(y)\sn^{2p+1}(y)}\,.
\eea

\item Starting from these seven solutions, we can generate further new
solutions by considering the transformations $y \rightarrow y+K(m)$,
$y \rightarrow y+iK'(m)$, $y \rightarrow y+K(m)+iK'(m)$ and 
interchanging $a,b,f,g$ appropriately. One can
show that in this way, by starting from the solution (\ref{16}) with
$b=1/2$ one has 20 independent solutions.

\item There is one more remarkable symmetry associated with eqs.
(\ref{12}) and (\ref{13}). Note that eq. (\ref{13})
is invariant under $t \rightarrow -t$ followed by $i \rightarrow -i.$ 
Under this transformation, ansatz 
(\ref{12}) becomes
\be\label{12a}
\phi(y)=[\cn(y,m)-i\sn(y,m)]^{-t} Z(y)\,,
\ee
and hence it follows that solutions with the ansatz (\ref{12}) and
(\ref{12a}) as well as with the ansatz $\phi(y)=
[\cn(y,m)-i\sn(y,m)]^{t} Z(y)$  are degenerate in energy 
(i.e. have the same value of 
$R$). This then gives us 12 additional solutions.

\item  Thus starting with
the solution (\ref{16}), for $b=1/2$, we can generate 32 new sets of 
quasi-periodic solutions of Heun's equation. 
In particular, 8 of these solutions have $\gamma=\frac{1}{2}-p,
\frac{3}{2}+p, \delta=\frac{1}{2}+p-N, \frac{3}{2}+N-p, \epsilon=0,
2$; another 8 of these solutions have $\gamma=\frac{1}{2}+p-N,
\frac{3}{2}+N-p, \delta =\frac{1}{2}-p, \frac{3}{2}+p, \epsilon = 
1\pm t$, another 8 have $\gamma=0,2, \delta = 1 \pm t, \epsilon =
\frac{1}{2}-p, \frac{3}{2}+p$ and finally, another 8 have $\gamma=
1 \pm t, \delta = 0,2, \epsilon=\frac{1}{2}+p-N, \frac{3}{2}+N-p$.
The corresponding values of $\alpha,\beta$ are easily 
obtained by using $\alpha =\frac{a+\gamma+\delta+\epsilon-1/2}{2},
\beta=\frac{\gamma+\delta+\epsilon-a-3/2}{2}$ while $4mq$ 
can be computed by using eqs. (\ref{9})
and (\ref{16}).

\item Proceeding in the same way, by starting from the solution
(\ref{17}) with $b=3/2$, we also generate 32 independent solutions
in each of which $4mq$ takes two possible values. The corresponding
values of $\gamma,\delta,\epsilon$ are the same as in the $b=1/2$ case
except that the values $0,2$ are now replaced everywhere by
$-1,3$ respectively. 

\end{enumerate}

{\bf Case 2: $f$ half-integral}

We start from eq. (\ref{7}) and substitute the ansatz
\be\label{3.15}
\phi(y) = [\dn(y,m)+ik\sn(y,m)]^{t} Z(y)\,,
\ee
where $t$ is any real number and $k = \sqrt{m}$. 
It follows that $Z(y)$ satisfies the equation
\bea\label{3.16}
&&Z''(y)+[2ikt\cn(y,m)+2mb\frac{\sn(y,m)\cn(y,m)}{\dn(y,m)}
-2g\frac{\cn(y,m)\dn(y,m)}{\sn(y,m)}+2f\frac{\dn(y,m)\sn(y,m)}
{\cn(y,m)}]Z'(y) \nonumber \\
&&+[-(R+m t^2)+(Q-t^2)m\sn^2(y,m)-2itkg\frac{\dn(y,m)}{\sn(y,m)}
-2iktb(1-m)\frac{\sn(y,m)}{\dn(y,m)} \nonumber \\
&&+ikt(2b+2f+2g-1)\sn(y,m)\dn(y,m)]Z(y)=0\,,
\eea 
where $R$ and $Q$ are as given by eq. (\ref{8}).
Not surprisingly, $Z(y)=$ constant is a solution with energy
$E=(4mt^2+1)/4$ provided $b=g=0,~f=1/2,~a=t-1/2$ (i.e. $b+f+g=1/2$).  

One can build solutions for higher values of $b+f+g$ from here. In
particular, for $b+f+g=2M+1/2$, we consider the ansatz
($M=0,1,2,...$)
\be\label{3.17}
Z(y)=\sum_{k=0}^{M} A_k\, \sn^{2k}
(y,m)+\sn(y,m)\dn(y,m)\sum_{k=0}^{M-1} B_k\, \sn^{2k} (y,m)\,,
\ee
while if $b+f+g=2M+3/2$ we take the ansatz ($M=0,1,2,...$)
\be\label{3.18}
Z(y)=\dn(y,m)\sum_{k=0}^{M} A_k\, \sn^{2k}
(y,m)+\sn(y,m)\sum_{k=0}^{M} B_k\, \sn^{2k} (y,m)\,.
\ee
Substitution into eq. (\ref{3.16}) yields analytic expressions for the
energy eigenvalues and eigenfunctions for arbitrary $M$ for 
$f=1/2$ and $f=3/2$. In particular, for $f=1/2$, we find that
\be\label{3.19}
f=1/2\,,~ g=p\,,~b+g=N\,,~ a=t-1/2\,,~~E=[mt^2+(g+f)^2]\,,
\ee
where both $b,g$ are nonnegative integers satisfying $b+g=N$ with
$N=0,1,2,...$ .

Similarly, when $f=3/2,a=t-1/2,g=p,b+g=N$ we find that
\be\label{3.20}
E=[(1+t^2)m+(g+f)^2] -(2g+1) \pm \sqrt{(2g+1)^2
+4m(N+1)(b-g)-4m(1-m)t^2}\,
\ee
where, $b$ and $g$ are again nonnegative integers. In all these cases, the
corresponding eigenfunctions have the form as given above in eqs.
(\ref{3.17}) and (\ref{3.18}). 
For small values of $N$, the explicit coefficients $A_k,B_k$ in the
eigenfunction expressions can be easily written down. 

Using eq. (\ref{8}), we can 
write down the corresponding solutions of Heun's eq. (\ref{2}).
They are given by
\be\label{3.21}
\gamma = \frac{1}{2}-p\,,~\delta=0\,,~\epsilon=\frac{1}{2}-N+p\,,~
\alpha=-\frac{N-t}{2}\,,~ \beta=-\frac{N+t}{2}\,,~4mq=m(N^2-t^2)\,.
\ee
\bea\label{3.22}
&&\gamma = \frac{1}{2}-p\,,~\delta =-1\,,~\epsilon=\frac{1}{2}-N+p\,,~
\alpha=-\frac{N+1-t}{2}\,,~ \beta=-\frac{N+1+t}{2}\,, \nonumber \\
&&4mq=m(N^2-t^2-1)+(2p+1) \pm \sqrt{(2p+1)^2 
+4m(N+1)(N-2p)-4m(1-m)t^2}\,.
\eea
Corresponding to each of these two solutions, we can again
write down 32 independent sets of solutions as shown above in Case 1 when $b$ was
half-integral.

{\bf Case 3: $g$ half-integral}

We start from eq. (\ref{7}) and substitute the ansatz
\be\label{3.23}
\phi(y) = [\dn(y,m)+k\cn(y,m)]^{t} Z(y)\,,
\ee
where $t$ is any real number. It then follows that $Z(y)$ satisfies the equation
\bea\label{3.24}
&&Z''(y)+[-2kt\sn(y,m)+2mb\frac{\sn(y,m)\cn(y,m)}{\dn(y,m)}
-2g\frac{\cn(y,m)\dn(y,m)}{\sn(y,m)}+2f\frac{\dn(y,m)\sn(y,m)}
{\cn(y,m)}]Z'(y) \nonumber \\
&&+[-R+(Q+t^2)m\sn^2(y,m)-2ktb\frac{cn(y,m)}{\dn(y,m)}
-2ktf\frac{\dn(y,m)}{\cn(y,m)} \nonumber \\
&&+kt(2b+2f+2g-1)\cn(y,m)\dn(y,m)]Z(y)=0\,,
\eea 
where $R$ and $Q$ are as given by eq. (\ref{8}).
Clearly, $Z(y)=$ constant is a solution with energy
$E=(1+m)/4$ provided $b=f=0,~g=1/2,~a=t-1/2$ (i.e. $b+f+g=1/2$).  

One can build solutions for higher values of $b+f+g$ from here. In
particular, when $b+f+g=2M+1/2$, we consider the ansatz
($M=0,1,2,...$)
\be\label{3.25}
Z(y)=\sum_{k=0}^{M} A_k \sn^{2k}
(y,m)+\cn(y,m)\dn(y,m)\sum_{k=0}^{M-1} B_k \sn^{2k} (y,m)\,,
\ee
while if $b+f+g=2M+3/2$ then we consider the ansatz ($M=0,1,2,...$)
\be\label{3.26}
Z(y)=\cn(y,m)\sum_{k=0}^{M} A_k \sn^{2k}
(y,m)+\dn(y,m)\sum_{k=0}^{M} B_k \sn^{2k} (y,m)\,.
\ee
Substitution into eq. (\ref{3.24}) leads to analytic expressions for the
energy eigenvalues and eigenfunctions for arbitrary $M$ when
$b=1/2, 3/2$. In particular, for $b=1/2$, we find that
\be\label{3.27}
g=1/2\,,~ b+f=N\,,~ a=t-1/2\,,~~E=[(f+g)^2+m(g+b)^2]\,,
\ee
where both $b,f$ are nonnegative integers satisfying $b+f=N$ with
$N=0,1,2,...$.
Similarly, when $g=3/2,a=t-1/2,b+f=N$ we obtain
\be\label{3.28}
E=(f+g)^2+m(g+b)^2-[1+2f+(2b+1)m] 
\pm \sqrt{(1-m)[(2f+1)^{2}-(2b+1)^{2}
m]+4mt^2}\,.
\ee
In all these cases, the
corresponding eigenfunctions have the form given above in eqs.
(\ref{3.25}) and (\ref{3.26}).

We now write down the solutions of Heun's eq. (\ref{2}) corresponding
to solutions (\ref{3.27}) and (\ref{3.28}):
\be\label{3.29}
\gamma=0\,,~\delta=\frac{1}{2}-N+p\,,~\epsilon = \frac{1}{2}-p\,,~
\alpha=-\frac{N-t}{2}\,,~ \beta=-\frac{N+t}{2}\,,~4mq=0\,;
\ee
\bea\label{3.30}
&&\gamma=-1\,,~\delta=\frac{1}{2}-N+p\,,~\epsilon = \frac{1}{2}-p\,,~
\alpha=-\frac{N+1-t}{2}\,,~ \beta=-\frac{N+1+t}{2}\,, \nonumber \\
&&4mq=2(N-p)+(2p+1)m \pm \sqrt{(1-m)[(2N-2p+1)^2-(2p+1)^2 m]+4mt^2}\,. 
\eea
Corresponding to each of these two solutions, we can again
write down 32 independent sets of solutions as done above when $b$ was
half-integral.

Summarizing, in this letter we have obtained new 192 sets of
quasi-periodic solutions of Heun's eq. (\ref{2}), each of which is doubly
degenerate. In each set, solutions exist with polynomials of arbitrary
order $N$, and for each $N$, there are $N+1$ distinct solutions. 

We would like to thank K. Takemura for informing us about some related work 
and recent developments in the mathematics literature on GAL potentials 
and Heun's equation.



\begin{thebibliography}{99}

\bibitem{ron} For an excellent up to date mathematical summary of
Heun's equation, see A. Ronveaux (ed.), {\it Heun's
Differential Equation} (Oxford University Press, 1995).
\bibitem{mai} R.S. Maier, arXiv:math.CA/0408317.
\bibitem{erd} A. Erd\'elyi et al. (ed.), {\it Higher Transcendental
Functions} (Bateman Manuscript Project) Vol. III (McGraw-Hill, 1955).
\bibitem{jnk} N.H. Christ and T.D. Lee, Phys. Rev. {\bf D12} (1975)
1606; D.P. Jatkar, C.N. Kumar and A. Khare, Phys. Lett. {\bf
A142} (1989) 200; A. Khare and B.P. Mandal, Phys. Lett. {\bf A239} 
(1998) 197.
\bibitem{bre} 
Y. Brihaye, S. Giller, P. Kosinski and J. Kunz, Phys. Lett. {\bf B293}
(1992) 383; S. Briabant and Y. Brihaye, Jour. Math. Phys. {\bf 34} (1994)
2107; 
Y. Brihaye, S. Giller and P. Kosinski, Jour. Phys. {\bf
A28} (1995) 421.
\bibitem{tak} K. Takemura, Comm. Math. Phys. {\bf 235} (2003) 467; Jour. 
Nonlinear Math. Phys. {\bf 11} (2004) 21; arXiv:math.CA/0406141.
\bibitem{bklms} R.K. Bhaduri, A. Khare, J. Law, M.V.N. Murthy and
D. Sen, Jour. Phys. {\bf A30} (1997) 2557.
\bibitem{stu} M. Suzuki, E. Takasugi and H. Umetsu, Prog. Theor. Phys.
{\bf 100} (1998) 491.
\bibitem{dst} P. Dorey, J. Suzuki and R. Tateo, Jour. Phys. {\bf A37}
(2004) 2047.
\bibitem{abr} For the properties of Jacobi elliptic functions, see,
for example, M. Abramowitz and I. Stegun,
{\it Handbook of Mathematical Functions} (Dover, 1964);
I. S. Gradshteyn and I. M. Ryzhik, 
{\it Table of Integrals, Series and Products} (Academic Press, 2000).
\bibitem{ks1} A. Khare and U. Sukhatme,
Jour. Math. Phys. {\bf 40} (1999) 5473.
\bibitem{ks2} A. Khare and U. Sukhatme,
Jour. Math. Phys. {\bf 42} (2001) 5652.
\bibitem{tv} A. Treibich and J.-L. Verdier, C.R. Acad. Sci. Paris {\bf 311} (1990) 51. 
\bibitem{ks3} A. Khare and U. Sukhatme,
arXiv:math-ph/0505027.
\bibitem{gp} For a recent attempt, see for example, N. Gurappa and
P.K. Panigrahi, Jour. Phys. {\bf A37} (2004) L605.
\end{thebibliography}
\end{document}